# Thoughts On Incorporating HPRF In A Linear Cooling Channel

Juan C. Gallardo[a] and Michael S. Zisman[b]

[a]Brookhaven National Laboratory, Upton, NY 11973
[b]Lawrence Berkeley National Laboratory, Berkeley, CA 94720

**Abstract.** We discuss a possible implementation of high-pressure gas-filled RF (HPRF) cavities in a linear cooling channel for muons and some of the technical issues that must be dealt with. The approach we describe is a hybrid approach that uses high-pressure hydrogen gas to avoid cavity breakdown, along with discrete LiH absorbers to provide the majority of the energy loss. Initial simulations show that the channel performs as well as the original vacuum RF channel while potentially avoiding the degradation in RF gradient associated with the strong magnetic field in the cooling channel.

## INTRODUCTION

There is by now considerable experimental evidence that the maximum gradient in a normal conducting vacuum RF cavity degrades markedly when the cavity is immersed in a strong axial magnetic field. Unfortunately, this is precisely the configuration proposed [1–3] for ionization cooling of muons. Recent studies [4] using a cavity filled with high-pressure hydrogen gas show no degradation in maximum gradient in this configuration. For this reason, it seems prudent to begin investigating the technical aspects of implementing HPRF in the Study 2a [2] cooling channel.

## SAFETY APPROACH

The safety approach we follow here is that developed in designing the MICE experiment [5]. In simple terms, the primary objective is to deliver a system that is both safe and usable. We note in passing that providing either requirement by itself is much easier than providing both simultaneously.

Focusing specifically on hydrogen safety, the MICE experimental design was required to *i)* maintain separation of hydrogen and oxygen atmospheres, and *ii)* avoid ignition sources in contact with hydrogen. These are redundant requirements, in the sense that either suffices to prevent an unsafe condition. Clearly, requirement *ii)* cannot be met in the case of an HPRF system. The best one can do is to ensure that the RF power cannot be on unless the cavities are properly pressurized.

In MICE, we also insisted on a design that would tolerate two things going wrong simultaneously. Failures often occur in pairs, and this criterion makes the design considerably more robust from a safety standpoint.

From the institutional perspective, there are many rules to satisfy, including explosive atmosphere regulations, pressure vessel codes, and the like. Hydrogen is flammable and explosive in air over a very broad concentration range. In MICE, all vacuum vessels were designated as pressure vessels (which implies testing to 1.25 times the design operating pressure), there are two barriers between hydrogen and oxygen, and there are dedicated hydrogen evacuation paths for the experimental apparatus and the hydrogen storage area. Similar approaches will be needed for an HPRF channel.

## HPRF ISSUES

There are key differences between an HPRF cooling channel and a "standard" (vacuum RF) channel. In the HPRF channel, the energy loss is distributed throughout the channel, rather than localized at discrete absorber locations. Moreover, the loss medium is gaseous rather than liquid or solid, which makes containment more of a challenge. This latter feature probably leads to a requirement—or at least a desire—for more modularity in the system design. While the addition of high-pressure hydrogen to the RF cavity likely increases the maximum allowable gradient [4], it is difficult to increase the energy loss correspondingly in a gas-filled system.

## STUDY 2A CHANNEL WITH HPRF

The Study 2a cooling channel [2] included 53 1.5-m cells (see Fig. 1), each with four 1-cm-thick LiH absorbers, two 201-MHz RF cavities, and a pair of opposite polarity superconducting solenoids. A gas-filled version of this channel (with the LiH windows removed) was studied by Fernow and Gallardo [6] several years ago. In the initial attempt, a channel having 124 atm of gaseous $H_2$ ($GH_2$) was used to produce the same energy loss that took place in the four LiH windows. This led to worse performance than the original Study 2a channel in the sense that fewer muons were produced within the acceptance of the downstream systems. An attempt at further optimization was ultimately successful. As shown in Fig. 2, taken from Ref. [6], at a gas pressure of about 200 atm, the channel recovered its original throughput, and at 250 atm, the performance was actually somewhat better than the original channel.

### Comments on Previous HPRF Study

It is not obvious that replacing *all* of the LiH absorbers with gaseous $H_2$ is the best approach. It requires ~200 atm of gas at room temperature, which gives rise to substantial engineering challenges. In particular, the isolation windows would have to be rather thick to withstand 200 atm. It is not clear how many isolation windows would actually be needed. The minimum number of isolation windows would be two, but this results in a high pressure hydrogen gas containment vessel on the order of 100 m in length. More segmentation may well be needed for safety reasons. Cooling the gas to liquid-nitrogen temperature would reduce the pressure needed by roughly a factor of four, but this is still a rather high pressure and the cooling brings its own engineering challenges.

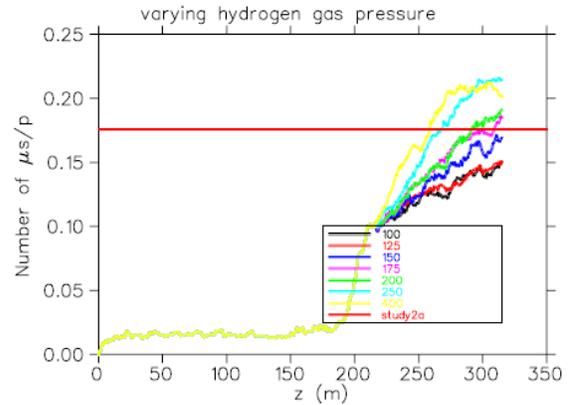

**FIGURE 2.** Throughput of the gas-filled Study 2a channel, measured by the number of muons per proton within the downstream acceptance, for various values (in atm) of $H_2$ gas pressure. The line at 0.17 μ/p represents the original Study 2a value. Taken from Ref. [6].

While the high pressure might well provide cavities that could operate at much higher gradients, this is likely to be vast "overkill" compared with the energy loss that can be achieved in gaseous hydrogen. Reducing the number of cavities mitigates the problem, e.g., halving the number of cavities with double the gradient proposed for Study 2a. Even so, it is probably impractical to feed the required power into an individual cavity. If the operating gradient is limited only by surface breakdown, a peak gradient of 50–60 MV/m might be possible. Reaching such a gradient in a 201-MHz cavity, however, would require ten times the input power of the nominal Study 2a cavity, implying some 45 MW of RF power into each cavity. Even doubling the gradient would require about 18 MW input power per cavity.

## ALTERNATIVE STRATEGY

We propose here an alternative, "hybrid" approach to converting the Study 2a channel to an HPRF version. Because the primary purpose of using HPRF is to avoid degradation of the cavity gradient due to the superimposed magnetic field, we use only enough gas to accomplish this task for the nominal Study 2a operating gradient of 15 MV/m. This requires much lower pressure than that needed to reach the material limit. Based on the measurements in [4] at 805 MHz, we expect that a pressure of 34 atm (at room temperature) will suffice. Eventually this parameter would need to be confirmed in tests at 201 MHz.

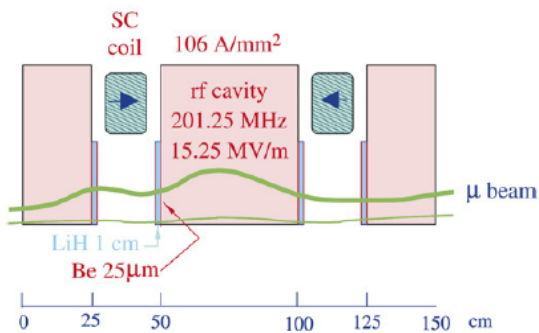

**FIGURE 1.** Schematic of Study 2a cooling channel.

At our specified pressure, the energy loss in the $GH_2$ is about one-quarter of that in the four original 1-cm-thick LiH windows. To compensate for the loss in the gas, we reduce the thickness of the LiH windows by 25%, to 0.75 cm. This is unlikely to be exactly correct, as there is some effect from the beta weighting of the energy loss, but it should be a reasonable starting point for optimizing channel performance.

This is the hybrid channel we study in this paper.

## EVALUATION OF HYBRID CHANNEL

To evaluate the performance of the proposed hybrid channel, we carried out ICOOL [7] simulations. Figure 3 shows a comparison of the transverse emittance cooling performance of the hybrid channel, using 34 atm of $GH_2$ along with four 0.75-cm LiH absorbers per cell, compared with the original Study 2a vacuum channel. Figure 4 shows the comparison in terms of throughput, that is, the number of muons per incident proton (within the downstream longitudinal and transverse acceptances of 150 mm and 30 π mm-rad, respectively) that reach the end of the channel. For both parameters, the comparison is quite favorable. In Fig. 3, we see that the final transverse emittance reached is essentially the same in both cases, despite the initial "spike" in emittance that results from passing through the thick (1-cm Ti) isolation window. In Fig. 4, the throughput for the hybrid channel was 10% better than that of the original vacuum channel.

To assess the effects of adding more isolation windows, we placed a third window at the center of the cooling channel and repeated the simulation. These results are presented in Figs. 5 and 6. The effect of even a single extra window is clearly visible, and causes some degradation in performance.

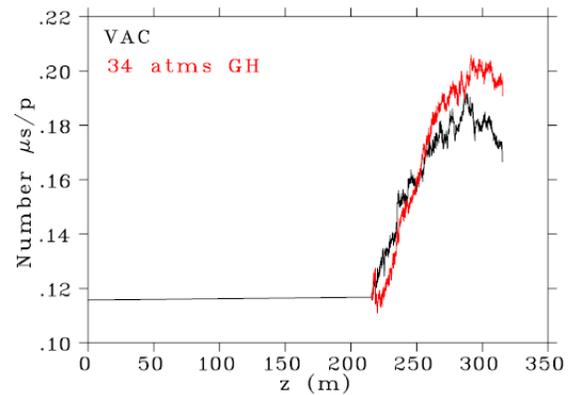

**FIGURE 4.** Simulated throughput of the hybrid cooling channel (red line) compared with that of the original Study 2a channel (black line).

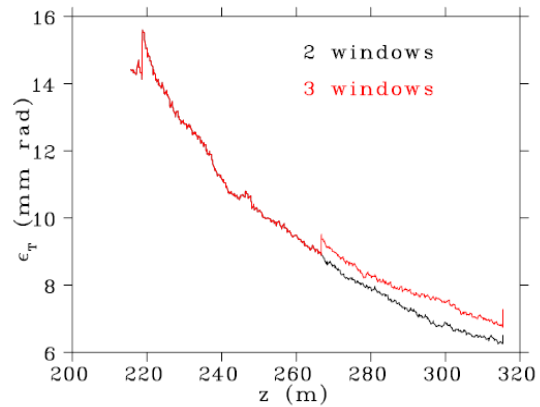

**FIGURE 5.** Simulated transverse cooling performance of hybrid cooling channel with two isolation windows (black line) and with three windows (red line). The "extra" spike in emittance at about 265 m is due to passage through the central Ti isolation window.

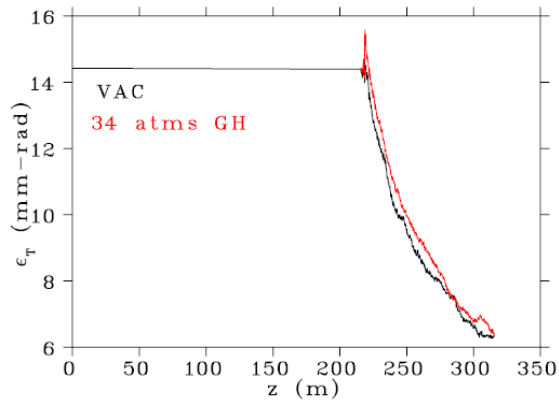

**FIGURE 3.** Simulated transverse cooling performance of hybrid cooling channel (red line) compared with original Study 2a channel (black line). The spike in emittance for the hybrid case is due to passage through a 1-cm-thick Ti isolation window at the entrance to the cooling channel.

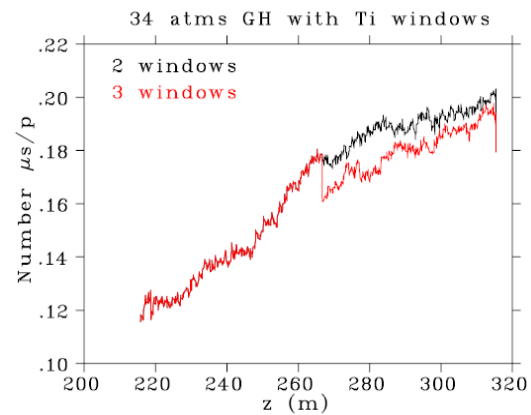

**FIGURE 6.** Simulated throughput for hybrid channel with two isolation windows (black line) and three isolation windows (red line). The particle loss at 265 m is due to scattering in the central Ti isolation window.

Given the observed degradation in performance due to even a single extra window, the need for such isolation must be examined carefully. Maintenance issues could be handled simply by having gate valves at suitable locations, provided they can be built to seal against the 34 atm gas pressure no matter which side of the valve it is on. This approach, of course, does not degrade performance during operation when the valves are open. Whether safety considerations dictate more subdivisions remains to be assessed. Note that we have made no attempt here to optimize the isolation window material, though this will undoubtedly be helpful. Clearly a lower Z material is favored if its strength is suitable. Aluminum is certainly a candidate, as it is known to resist hydrogen embrittlement—a critical requirement in this application.

## Comments on Implementation

The implementation of a gas-filled cooling channel is likely to be challenging. As already alluded to, having a continuous 80–100 m pipe filled with high-pressure hydrogen will have safety implications and may not be acceptable. In this configuration, any problem encountered would involve the entire hydrogen inventory. Even routine removal of the gas for storage would require large storage volumes or very high pressure. The venting system would likely need to be distributed along the channel in any case to give adequate conductance. For these reasons, a modular system, with independent gas supplies, seems more desirable. Unfortunately, such a system comes with the obvious drawback of more isolation windows unless the use of gate valves proves to be acceptable.

Another issue of consequence is that of hydrogen embrittlement, which can weaken many materials. In a cooling channel, Cu, Be, and LiH are common materials whose behavior must be ascertained. The materials of the gas containment system—a pressure vessel—must particularly be certified for $GH_2$ use.

Although operating the channel at liquid-nitrogen temperature will reduce the pressure requirements, it greatly complicates the engineering of the channel. One must consider the means to cool the RF cavities, differential contraction of various materials, the need for an insulating vacuum, and the like. Cooling the gas also makes it difficult to add an electron scavenger, e.g., $SF_6$, to the gas mixture to improve its breakdown performance when subjected to an intense beam of ionizing particles. For these reasons, we presently favor operation at room temperature.

## POSSIBLE IMPLEMENTATION IDEAS

In order to get a preliminary look at some of the specific issues that might be involved in creating such a channel, we have briefly considered two possible scenarios, illustrated schematically in Figs. 7 and 8.

Figure 7 illustrates the case where the cavity and beam pipe contain the high-pressure gas, and these are surrounded by an insulating vacuum vessel. This configuration lends itself better to the possibility of cryogenic operation, though the engineering caveats mentioned earlier still apply. Another advantage of this approach is that, from a safety perspective, the "hydrogen zone" can be contained within the experimental apparatus.

In this implementation, the cavity walls and beam pipe walls must be strong enough to contain 34 atm of $GH_2$. On the other hand, the LiH windows cannot be subjected to this pressure so the system must be filled in such a way as to avoid creating a significant differential pressure at those locations. The RF input couplers must also be pressurized equally on both sides, unless the coupler can be made strong enough to handle the pressure. Moretti [8] has suggested an epoxy-filled RF window that appears strong enough to handle pressures in this range, and this would simplify the design somewhat if acceptable from both the safety and RF standpoints.

A storage system for the hydrogen is needed to accommodate the evacuation of the channel for maintenance. This could be either metal hydride beds or a large buffer tank. The former approach would be physically more compact, but reabsorbs gas relatively slowly. The latter approach is straightforward but takes up more space unless the gas is stored under very high pressure.

With the implementation illustrated in Fig. 8, the cavities and windows see no differential pressure if the system is filled carefully. In practice, one would try to size the pipes to preclude the buildup of differential pressure during either the filling or venting operation. The cavities and tuners could then be of a design similar, or possibly identical, to those used in MICE [5]. The standard bellows connections between modules would be a challenge, as the bellows would be part of the pressure containment system.

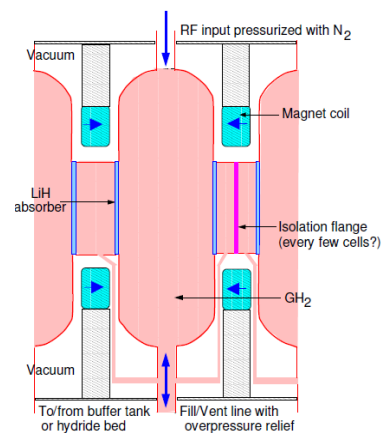

**FIGURE 7.** Schematic illustration of hybrid channel implementation with insulating vacuum vessel. The $GH_2$ (shaded area) fills only the cavity and beam pipe, but not the surrounding vacuum vessel.

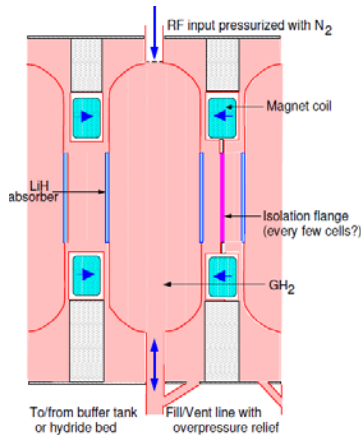

**FIGURE 8.** Schematic illustration of hybrid channel implementation without insulating vacuum vessel. The $GH_2$ (shaded area) fills the cavity, beam pipe, and surrounding containment vessel.

It is likely that this configuration would require that the entire accelerator housing in this region be classified as a hydrogen zone. This would require restrictions on all electronics devices in the area to make sure they are rated for operation in a hydrogen atmosphere. Special lighting, light switches, pumps, and diagnostics instrumentation would be needed, not all of which are commercially available.

Cryogenic operation in this configuration would require a vacuum-insulated outer layer. Even so, the ability to warm up individual sections would be a challenge. The vacuum-insulated layer would avoid the necessity to classify the surrounding area as a hydrogen zone, and thus might be an attractive option even if cryogenic operation were not envisioned.

## SUMMARY

We took an initial look at the implications of using HPRF in a linear cooling channel. A hybrid channel is proposed, wherein $GH_2$ provides protection against cavity breakdown and LiH absorbers are the primary energy loss medium. We believe that the parameters of the hybrid channel will be more easily realized than those of previous HPRF channels. Initial performance estimates are encouraging, although the influence of isolation windows remains a potential impediment to achieving good transmission.

The pros and cons of two possible implementation schemes were examined, both of which have challenging aspects. Issues associated with cryogenic operation of the channel were briefly considered. Our tentative conclusion for the hybrid channel is that the engineering challenges outweigh the advantage of having a lower operating pressure, and for this reason we favor room-temperature operation of the system.

There remains considerable work to do before a hybrid channel can be considered a validated cooling channel option. Foremost among these will be an optimization of the isolation windows to see if their deleterious effects can be reduced, and materials investigations on hydrogen embrittlement. Still, if it turns out that HPRF is a reliable way to avoid RF cavity breakdown in a magnetic field, we believe that the hybrid concept suggested here will be a promising approach to the cooling channel of a Neutrino Factory or Muon Collider.

## ACKNOWLEDGMENTS

We wish to thank Michael A. Green, Alfred Moretti and Steve Virostek for helpful discussions on implementation issues. This work was supported by the Director, Office of Science, Office of High Energy Physics, of the U.S. Department of Energy under Contract Nos. DE-AC02-05CH11231 (LBNL) and DE-AC02-98CH10886 (BNL).